\documentclass[aps,prd,twocolumn,showpacs,amsmath,amssymb]{revtex4}
\usepackage{graphicx}
\usepackage{bm}

\begin{document}

\preprint{CRA/1}

\title{Multiflavor and multiband observations of neutrinos from core collapse supernovae}

\author{I.~Taboada}
\email{ignacio.taboada@physics.gatech.edu}
\affiliation{
  School of Physics and Center for Relativistic Astrophysics\\
  Georgia Institute of Technology\\
  Atlanta, GA 30332. USA}

\date{today}

\begin{abstract}
It has been proposed that the gamma ray burst - supernova connection
may manifest itself in a significant fraction of core collapse
supernovae possessing mildly relativistic jets with wide opening
angles that do not break out of the stellar envelope. Neutrinos would
provide proof of the existence of these jets. In the present letter we
calculate the event rate of $\gtrsim$100~GeV neutrino-induced cascades
in km$^3$ detectors. We also calculate the event rate for
$\gtrsim$10~GeV neutrinos of all flavors with the DeepCore low energy
extension of IceCube. The added event rate significantly improves the
ability of km$^3$ detectors to search for these gamma-ray dark
bursts. For a core collapse supernova at 10 Mpc we find $\sim$4 events
expected in DeepCore and $\sim$6 neutrino-induced cascades in
IceCube/KM3Net. Observations at $\gtrsim$10~GeV are mostly sensitive
to the pion component of the neutrino production in the choked jet,
while the $\gtrsim$100~GeV depends on the kaon component. Finally we
discuss extensions of the on-going optical follow-up programs by
IceCube and Antares to include neutrinos of all flavors at
$\gtrsim$10~GeV and neutrino-induced cascades at $\gtrsim$100~GeV energies. 
\end{abstract}


\maketitle

\section{Introduction}

Long duration gamma-ray bursts (GRBs) and core collapse supernovae are
known to be correlated \cite{grb030329,Hjorth:2003jt}. MeV photons
that give rise to GRBs are thought to be produced by accelerated
electrons in internal shock of jets emitted by the GRB
progenitor. Many details about GRB phenomenology remains
uncertain. These GRB jets have very narrow opening angles and very
large Lorentz boost factors ($\Gamma \gtrsim 300$.) GRBs are one of
the plausible candidates sources for the highest energy cosmic rays
\cite{Vietri,WaxmanBahcall}  and as such would observable in
$\sim$~100~TeV neutrinos by km$^3$ detectors such as IceCube
\cite{IceCube} and KM3Net \cite{KM3Net}.

Only a very small fraction of core collapse supernovae result in
GRBs. It has been speculated that the rate of production of jets in core
collapse supernovae may be significantly higher than the rate of GRBs,
but that often these jets are choked within the supernovae. Evidence
for this hypothesis is the observed asymmetry in the explosion of core 
collapse supernovae \cite{AssymmetrySN2008D,Tanaka}. Very recently
evidence for mildly relativistic jets ($\Gamma\sim$~1) that did break
from the progenitor has been presented for type Ic supernovae 2007gr
and 2009bb \cite{sn2007gr,sn2009bb}. Supernovae with hidden jets, supernovae
with mildly relativistic jets that manage to break out and long GRBs could form a
continuum class of astronomical objects. Neutrinos would provide us
with information about these hidden jets and early neutrino
observations could also lead the way in finding and studying
supernovae with mildly relativistic jets that do break out. A model
has been proposed by Razzaque, M\'esz\'aros and Waxman (henceforth RMW)
\cite{RMW}. This model was extended by Ando and Beacom (henceforth AB)
to include kaon production \cite{AB}. The RMW/AB spectrum is very soft
leading to $\gtrsim$100~GeV neutrino observations in IceCube/KM3Net. A
model for neutrino production in reverse shocks for both choked and
successful relativistic jets associated with SN type Ib has also been
proposed \cite{horiuchi}. While we did not consider this latter model
in the present letter, the broad conclusions still apply.

A promising way to search for $\gtrsim$100~GeV neutrinos from core collapse 
supernovae is the optical follow-up \cite{Followup}. Neutrino
multiplets in directional and time coincidence trigger observations by
fast robotic telescopes. This method has the advantage of being able
to dramatically reduce the intrinsic atmospheric neutrino
background because the signal search window is only
$O(100\mathrm{s})$. A neutrino multiplet may not be significant on its
own because the accidental rate due to atmospheric neutrino background
is O(10/year), but the subsequent observation of the rising light-curve
of a supernova in spatial and temporal coincidence with a neutrino
multiplet would be highly significant. Optical follow-up programs are
already in operation by IceCube \cite{FollowupIceCube} and Antares
\cite{FollowupAntares}.

The objective of this letter is to calculate the event rate from core
collapse supernovae as observed by $\sim$10~GeV detectors like
DeepCore \cite{deepcore}. We propose the optical follow-up programs to
be enhanced so as to require at least one $\gtrsim$100~GeV $\nu_\mu$
(minimum requirement to achieve good sky localization) and several
$\gtrsim$10~GeV events. We also calculate the event rate on
$\gtrsim$~100 GeV neutrino-induced cascades. A most interesting case
is that of KM3Net, which promises good angular reconstruction of
cascades. We propose the optical follow-up programs to be enhanced so
as to require at least one $\sim$100~GeV $\nu_\mu$ and one
$\sim$100~GeV neutrino-induced cascade. The enhanced sensitivity of
these two new modes of observation will allow IceCube/KM3Net to
provide a more thorough test of the RMW/AB model.

\section{Neutrino production in choked jets}

The RMW/AB model supposes a choked jet with bulk Lorentz boost factor
$\Gamma_b \sim 3$ and an opening angle $\theta_j \sim 0.3$. The kinetic energy
of the jet is set to $E_j = 3\times10^{51}$~erg in analogy to
GRBs. Also similar to GRBs the time variability of the engine is set
to $t_v \sim 0.1$~s. Neutrinos are produced in p-p interactions via
both pions and kaons. The accelerated proton spectrum is assumed to be
$dN_p/dE = E^{-2}$. The density and energies involved are similar to
those of atmospheric neutrinos in which muons do not decay, but
instead are subject to radiative cooling. Thus in the case of
neutrinos generated by pion decay, the neutrino flavor flux ratio
$\phi_{\nu_e}$:$\phi_{\nu_mu}$:$\phi_{\nu_\tau}$ is 0:1:0. In the case
of kaons there is a small flux of $\nu_e$ due to $K^0_{\mathrm L}$
decay, but the AB paper does not include neutral kaon contribution. We
therefore also assume 0:1:0 as the flavor flux ratio. Taking into
account vacuum oscillations the expected flavor flux ratio at Earth is
0.2:0.4:0.4 for both pions and kaons. Note that vacuum oscillations
were not taken into account in previous calculations \cite{RMW,AB}, we
include them here as they are  critical to describe $\nu_e$ and
$\nu_\tau$ fluxes at Earth. All our numbers include vacuum
oscillations except where explicitly noted.

The mesons product of p-p interactions have the same initial spectrum
as the protons, but they are subject to hadronic and radiative
cooling. This results in two break energies above which the meson
spectrum is steeper. Neutrinos follow the energy spectrum of their
parent mesons. The neutrino flux resulting from pion and kaon
contributions can be described as a doubly broken power law:  

\begin{equation} 
\frac{dN_\nu}{dE} = A \left\{ \begin{array}{lr}
E^{-2} & E > E_\nu^{(1)} \\
E_\nu^{(1)} E^{-3} & E_\nu^{(1)} E < E_\nu^{(2)} \\
E_\nu^{(1)} E_\nu^{(2)} E^{-4} & E_\nu^{(2)} < E < E_{max} 
\end{array}
\right.
\end{equation}

$A$ is the flux normalization (for all flavors combined) such that for
pions(kaons) $dN_\nu/dE$ is $5\times10^{-2}$~GeV$^{-1}$.cm$^{-2}$ 
($5\times10^{-5}$~GeV$^{-1}$.cm$^{-2}$) at $E_\nu^{(1)}$. The energies 
$E_\nu^{(1)}$ and $E_\nu^{(2)}$ are the energies above which hadronic 
and radiative cooling become relevant respectively. For pions (kaons)
these energies are: $E_\nu^{(1)} = 30$~GeV (200~GeV)
and $E_\nu^{(2)} = 100$~GeV (20~TeV). The values above
correspond to a supernova at 10 Mpc with default choices of
parameters. See RMW\cite{RMW} and AB\cite{AB} for details.

\section{Effective Areas and Event rates in IceCube/KM3Net/DeepCore}
\label{sec:effarea}

\begin{figure*}
\includegraphics[width=0.3\textwidth]{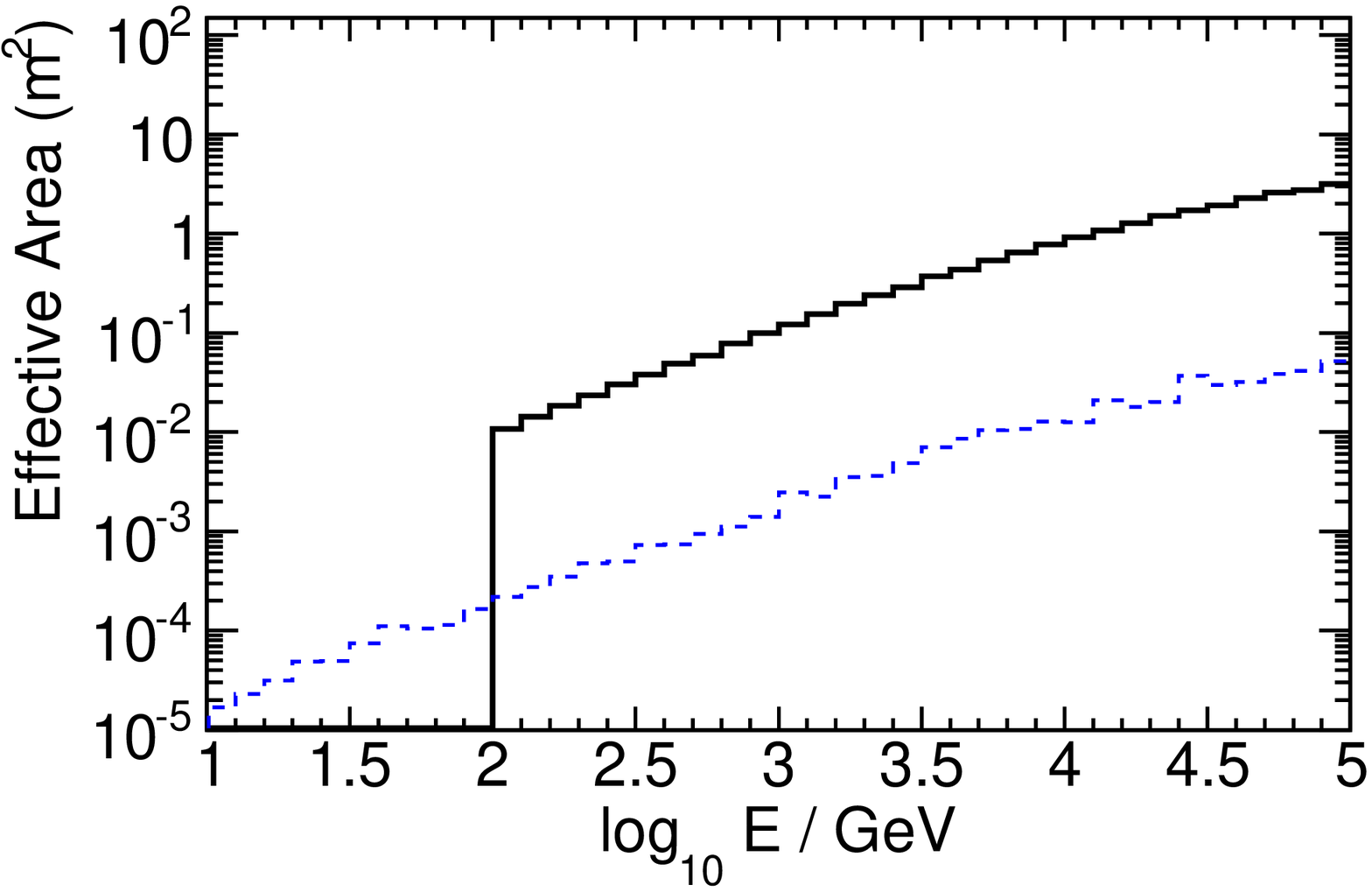}
\includegraphics[width=0.3\textwidth]{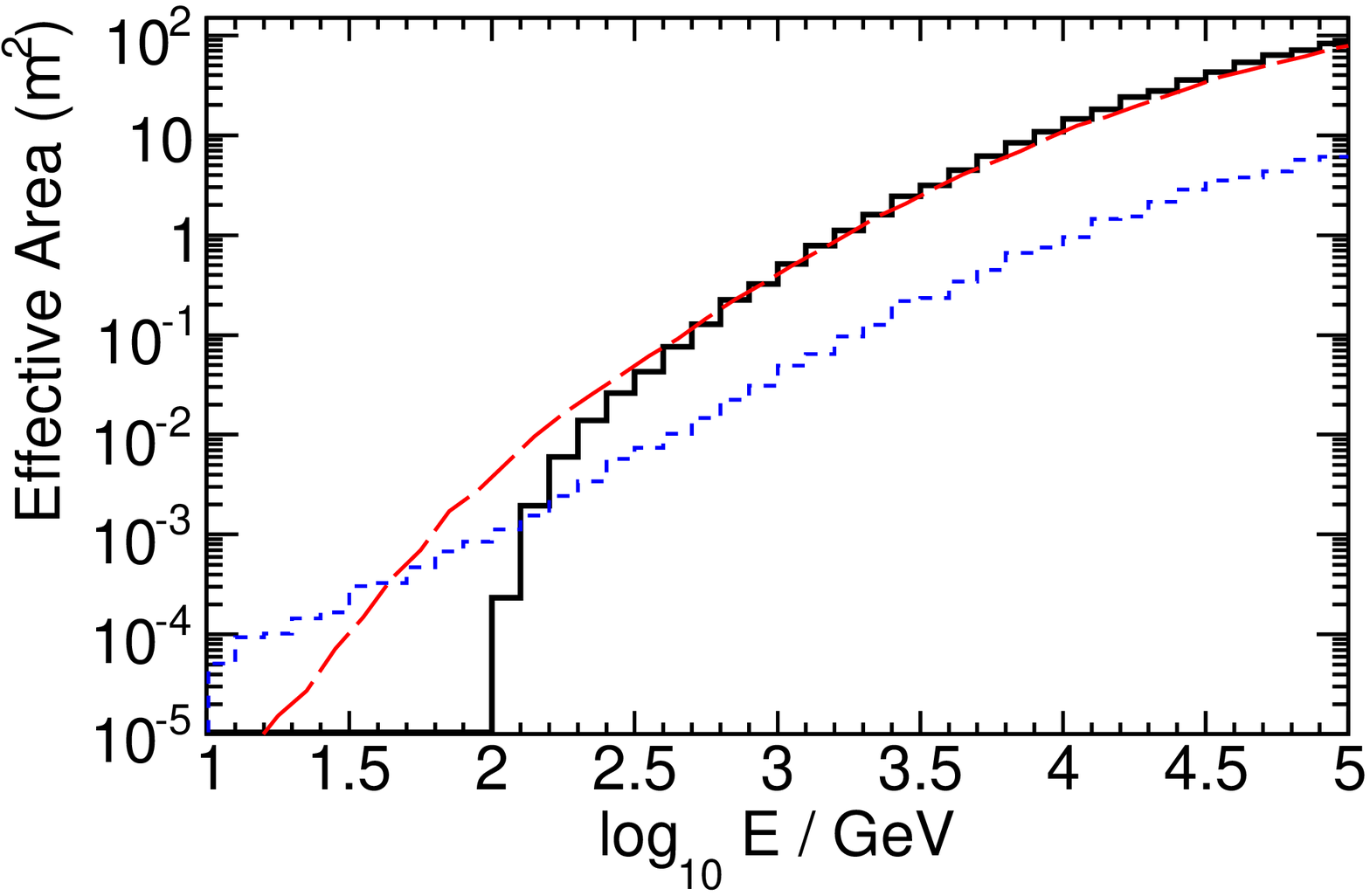}
\includegraphics[width=0.3\textwidth]{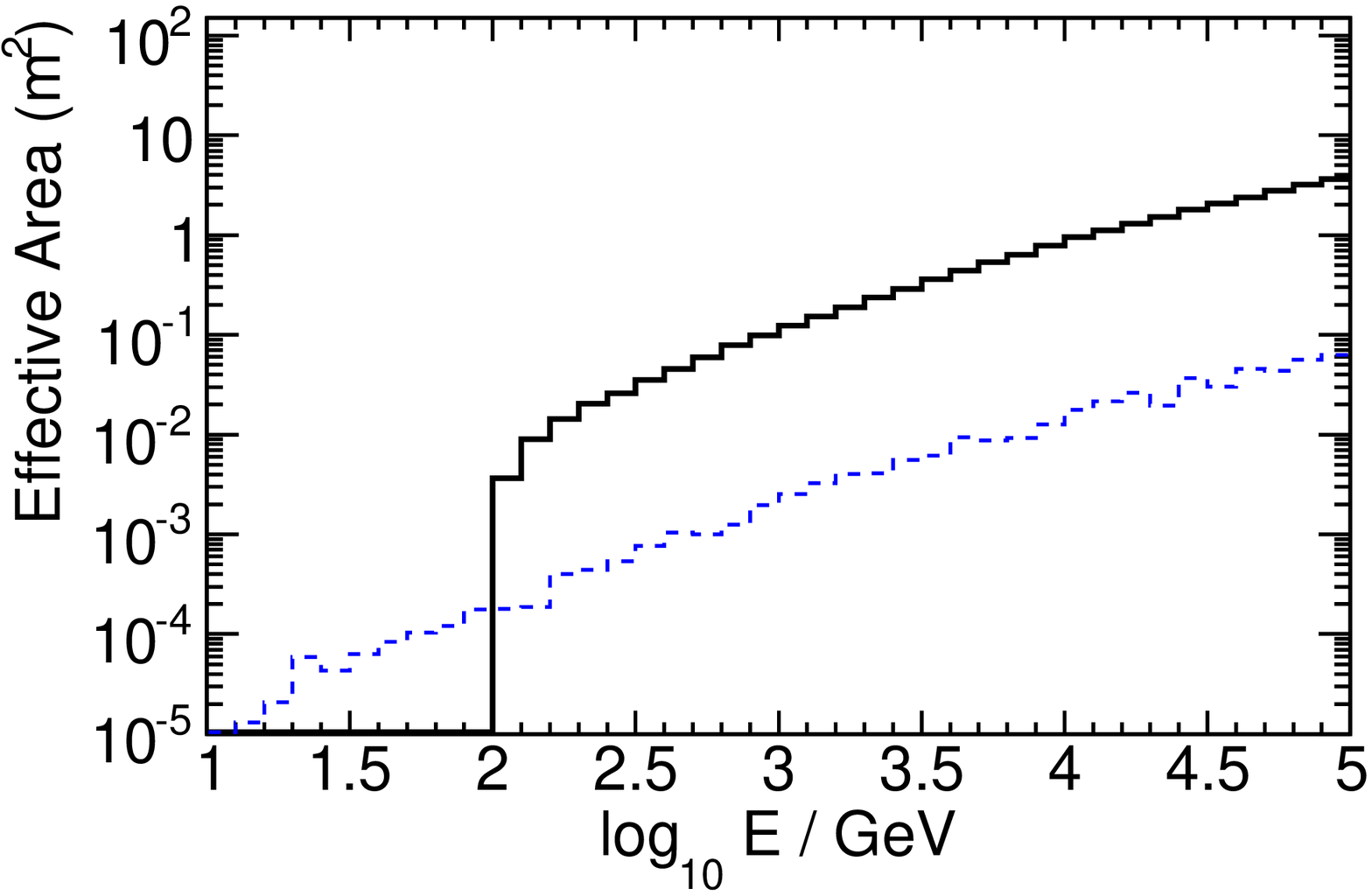}
\caption{\label{fig:Areas} The three panels show the neutrino
  effective areas for $\nu_e$ (left), $\nu_\mu$ (center) and $\nu_\tau$ (right). The effective areas shown are
  for IceCube (solid histogram - black online version) and DeepCore
  (short dashed histogram - blue online version) as calculated in this 
  letter. Also shown (long dashed curve - red online version) is the
  effective area for $\nu_\mu$ published by the IceCube
  collaboration. The effective area has been averaged over the
  northern hemisphere ($\delta$: $0^\circ$ to $90^\circ$). The effective
  area is also averaged over neutrinos and anti-neutrinos. The
  difference between IceCube's effective area for $nu_mu$ as
  calculated in this letter and as published by the IceCube
  collaboration is due to threshold effects. In this letter's
  calculation we have simplistically assumed that all muons with >
  100~GeV energy that go through IceCube are detected. For details see
  section \ref{sec:effarea}.} 
\end{figure*}

To calculate the expected number of events $N_{\mathrm obs}$ given a
neutrino flux $dN_\nu/dE$  the neutrino effective area $A_{eff}$ of
a detector is needed: 
\begin{equation}
\label{eq:effArea}
  N_{\mathrm{obs}} = \int dE A_{eff} (E) \frac{dN_\nu}{dE}
\end{equation}

The neutrino effective area of IceCube for muon neutrinos averaged
over the northern hemisphere and assuming equal $\nu$ and $\bar{\nu}$
fluxes has been published by the IceCube collaboration \cite{IceCubeEffectiveArea}
and it is reproduced in Fig.\ref{fig:Areas}. The neutrino effective
area of a detector can be calculated from first principles and the
procedure for doing so has been reported in many publications. In
particular \cite{HalzenGonzalez} and refs. therein have a careful
discussion of this subject.

The calculation of neutrino effective area requires knowledge of the
muon effective area, which in turn depends on detailed knowledge of
the operation of the detector. In the literature it is common to
assume that IceCube has an effective area of 1~km$^2$ for muons above
a threshold of 100~GeV. This approximation is a good one for the hard
spectra, such as that expected from the sources of cosmic rays. In
contrast this approximation is a bad one for soft fluxes, such as
those from the RMW/AB model. For RMW/AB most of the events observed
have energies that are close to the muon detection threshold and thus
detailed knowledge of the detector functioning (and hence the muon
effective area) are needed. As similar argument also applies to
DeepCore and cascade events.

We have used ANIS \cite{ANIS} to simulate neutrinos in the
vicinity of IceCube and DeepCore. We assume that IceCube is a cylinder
of 1000~m of height and 564~m in radius. The center of this cylinder
is placed 1950~m below the ice surface. DeepCore is simulated as a
cylinder of 350~m of height and 125~m of radius at 2275~m of depth
below the ice surface. The ANIS simulation takes into account the
ice/rock boundary, Earth neutrino absorption (though this is a very
small effect for the soft RMW/AB spectrum), neutral current
regeneration, etc. Our simulations include neutrinos of all
flavors. We have assumed flavor flux ratio as appropriate for an
the injected fluxes by either pions or kaons and we have taken into
account neutrino oscillations. We
have also assumed equal ratios of neutrino to anti-neutrino for a given
flavor. Muon propagation is calculated with MMC \cite{MMC}. We
assume that events are detected for the following conditions: a)
through-going muons or entering muons must have a minimum energy when
entering the detector cylinder b) the total visible energy (cascades,
muons, etc) of a contained event must be greater than the same
threshold as through-going events. For IceCube we have chose 100~GeV
as the threshold and for DeepCore 10~GeV. IceCube results also apply
to KM3Net. We show our calculations for neutrino effective areas for
IceCube and DeepCore in Fig. \ref{fig:Areas}.

Note good agreement between our calculations and IceCube published
effective for $\gtrsim$TeV. As described above, for low energies,
detailed simulation of the detector performance becomes important. For
DeepCore our simulation is in good agreement with the results
published by the IceCube collaboration \cite{deepcore}. The IceCube
collaboration has not published an effective area for neutrino-induced
cascades. Therefore our results concerning cascades will have the
largest uncertainty.

Performing the integral on Eq. \ref{eq:effArea} using IceCube's
published neutrino effective area with the RMW/AB flux for a core
collapse supernova at 10 Mpc confirms previous estimates of 
expected $\nu_\mu$ events \cite{AB}. Using IceCube' published
$\nu_\mu$ effective area we obtain an expectation of 11.2 events (28 
if we don't assume oscillations as AB did.) The comparison remains
equally valid at high energies for which threshold effects are less
relevant. Using our calculation of the $\nu_\mu$ effective area we
obtain 3.3 events. Our discrepancy with AB and the IceCube published
effective area can be attributed exclusively to threshold effects near
100~GeV.

\begin{figure*}
\includegraphics[width=0.3\textwidth]{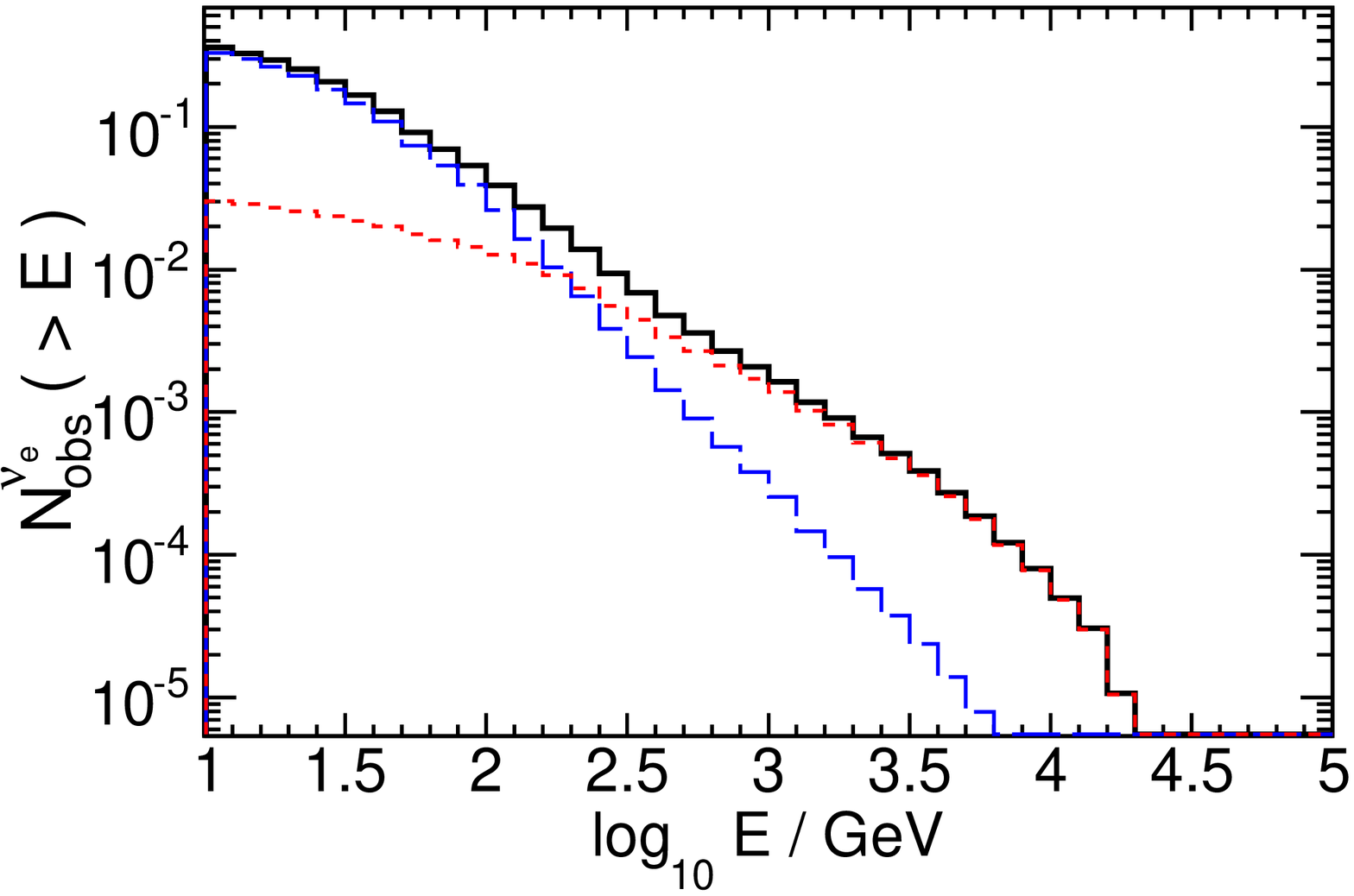}
\includegraphics[width=0.3\textwidth]{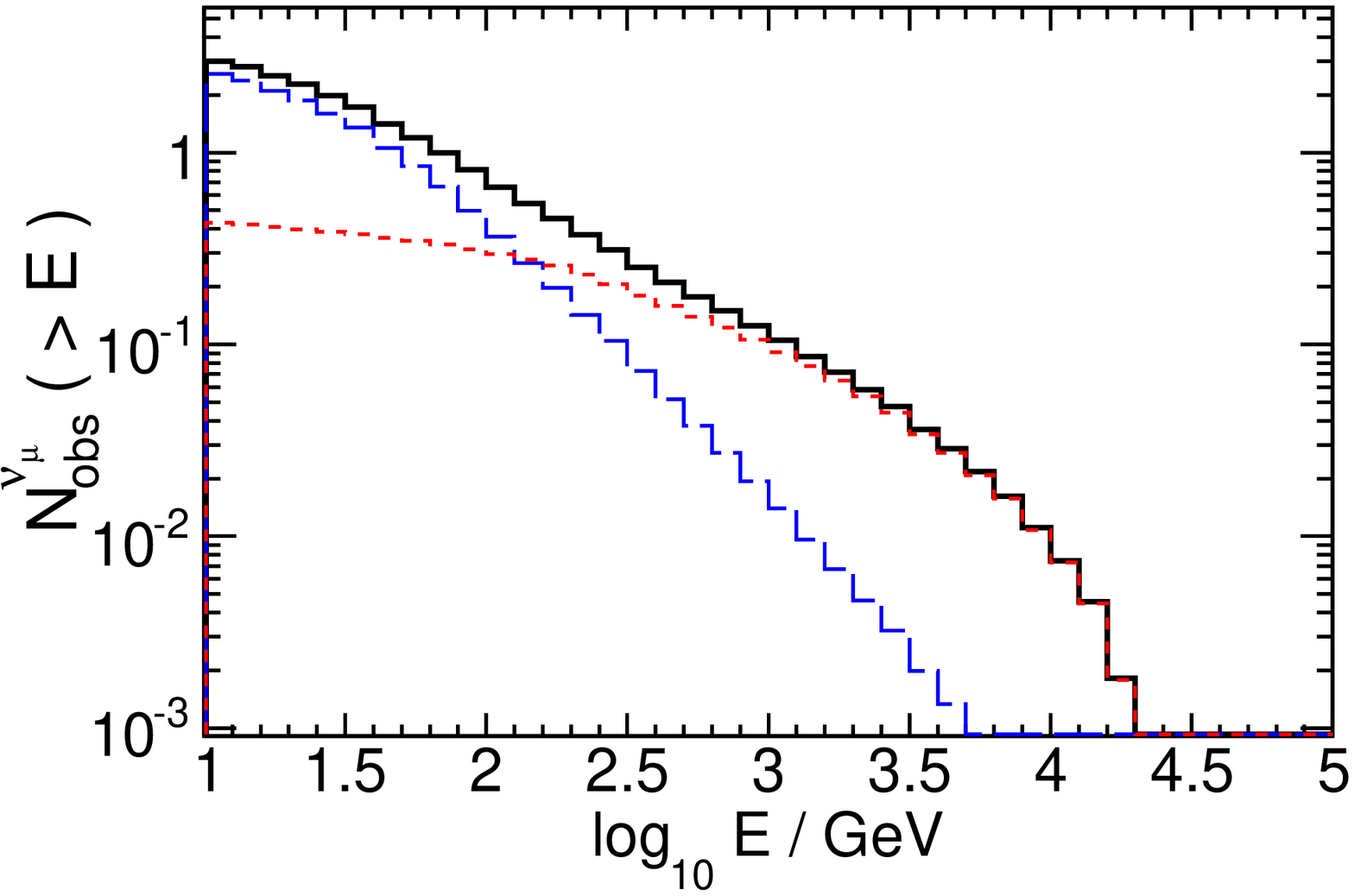}
\includegraphics[width=0.3\textwidth]{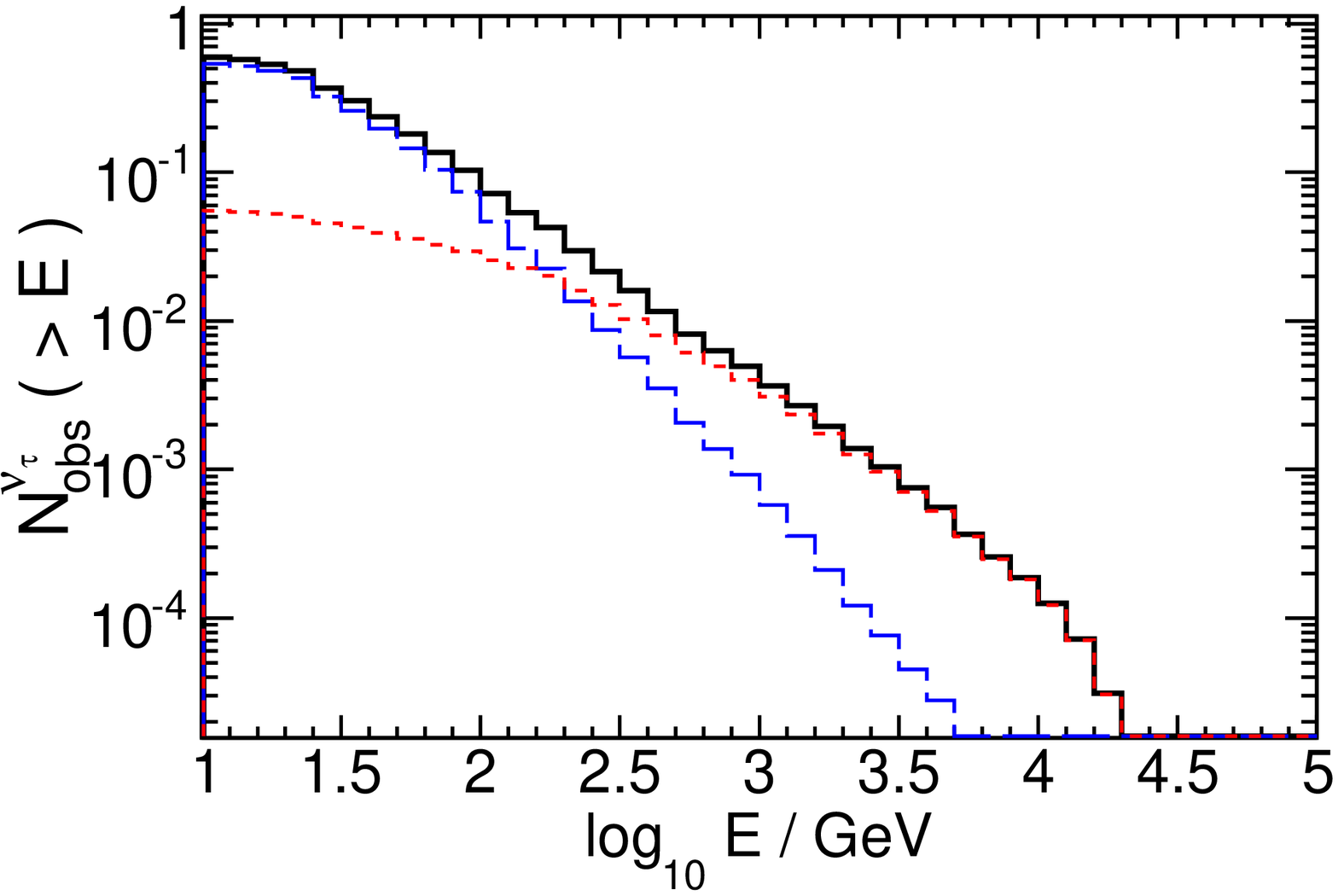}
\caption{\label{fig:DCrates} The panels show the expected signal
  above a given energy for DeepCore for $\nu_e$ (left), $\nu_\mu$
  (center) and $\nu_\tau$ (right). The solid (black in the online
  version) histogram is the total expectation of the RMW/AB model, the
  long dashed histogram (blue in the online version) is the pion 
  contribution and the short dashed (red in the online version)
  histogram is the kaon contribution.}
\end{figure*}

\begin{figure*}
\includegraphics[width=0.3\textwidth]{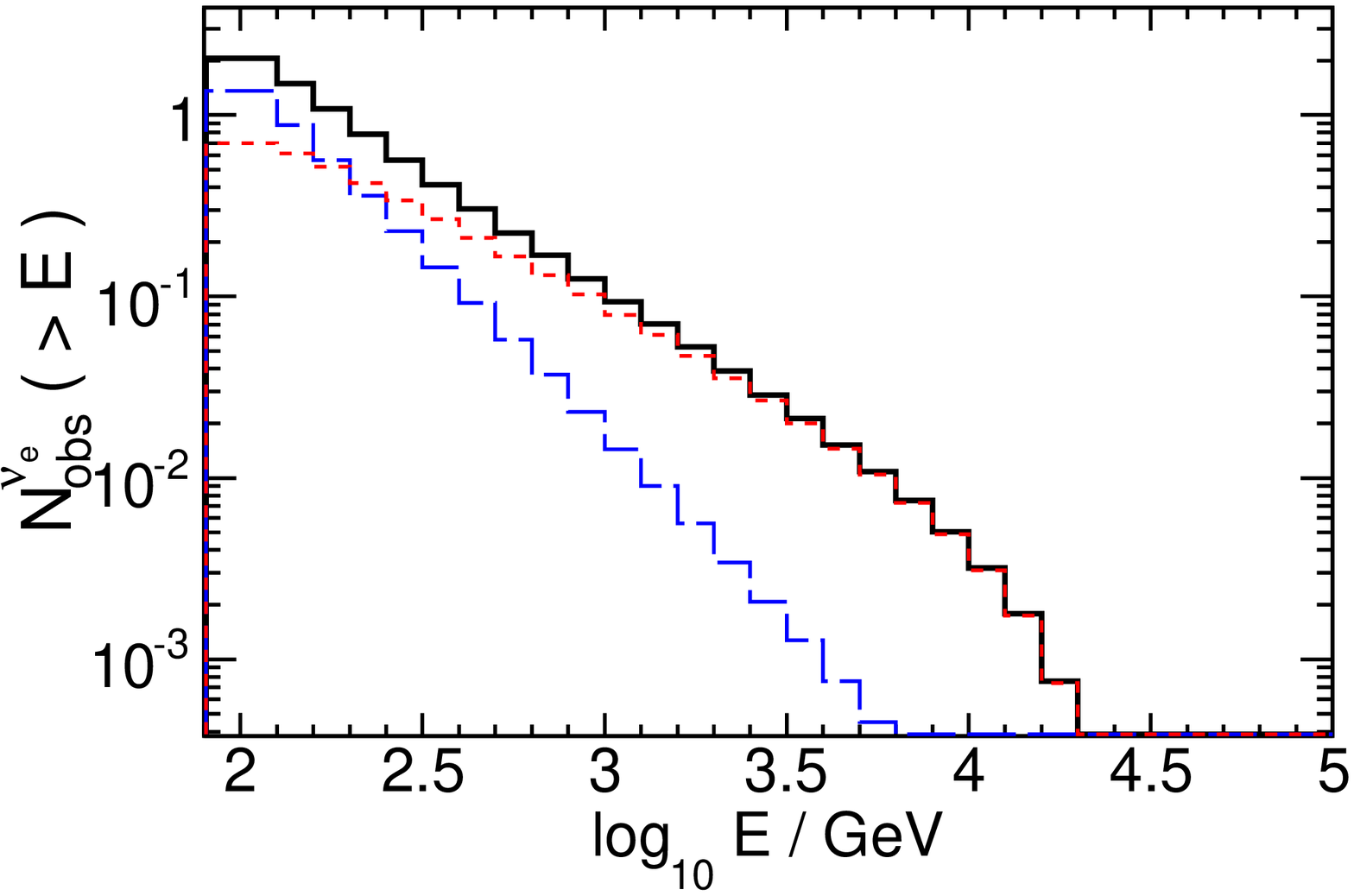}
\includegraphics[width=0.3\textwidth]{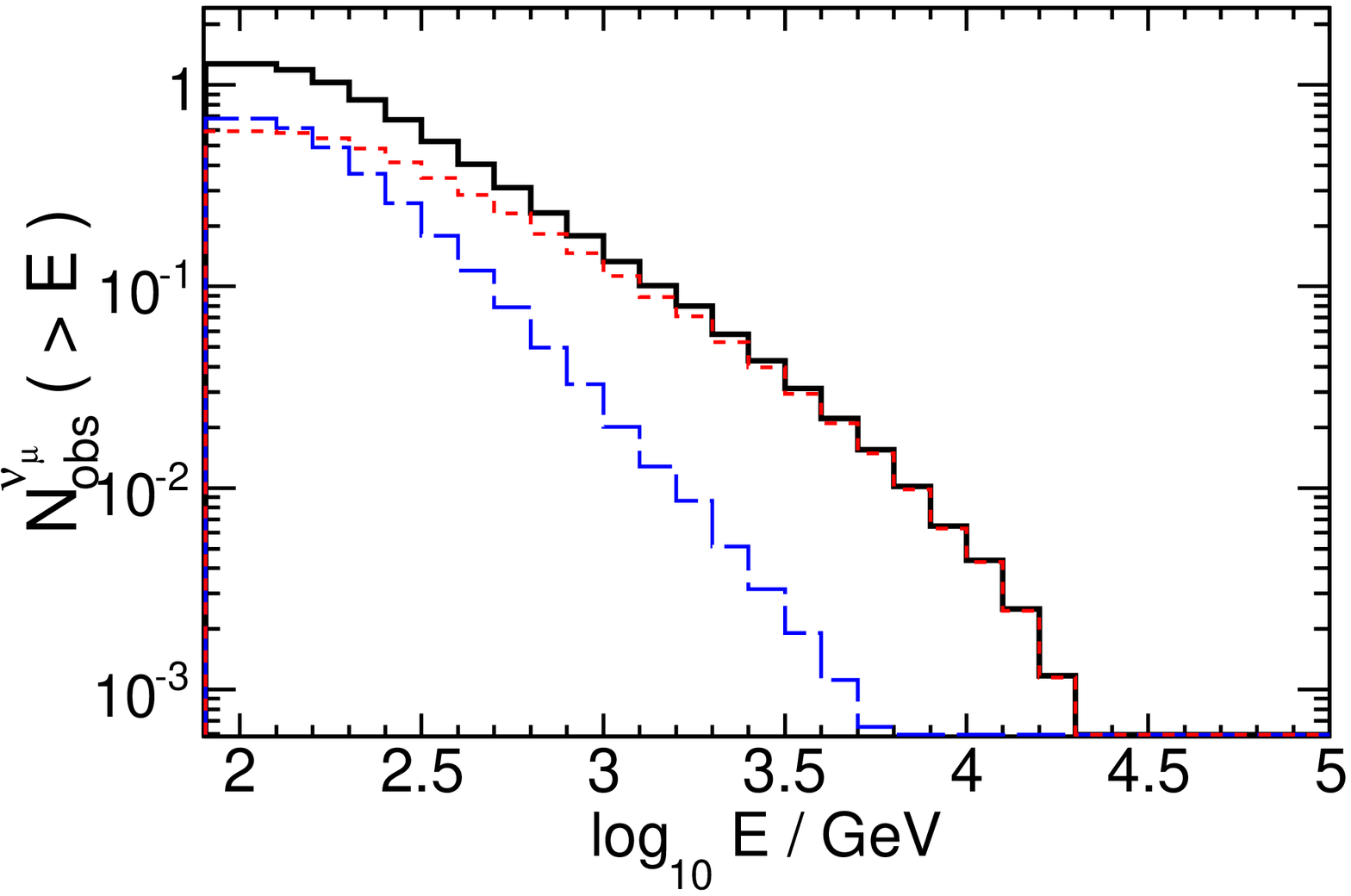}
\includegraphics[width=0.3\textwidth]{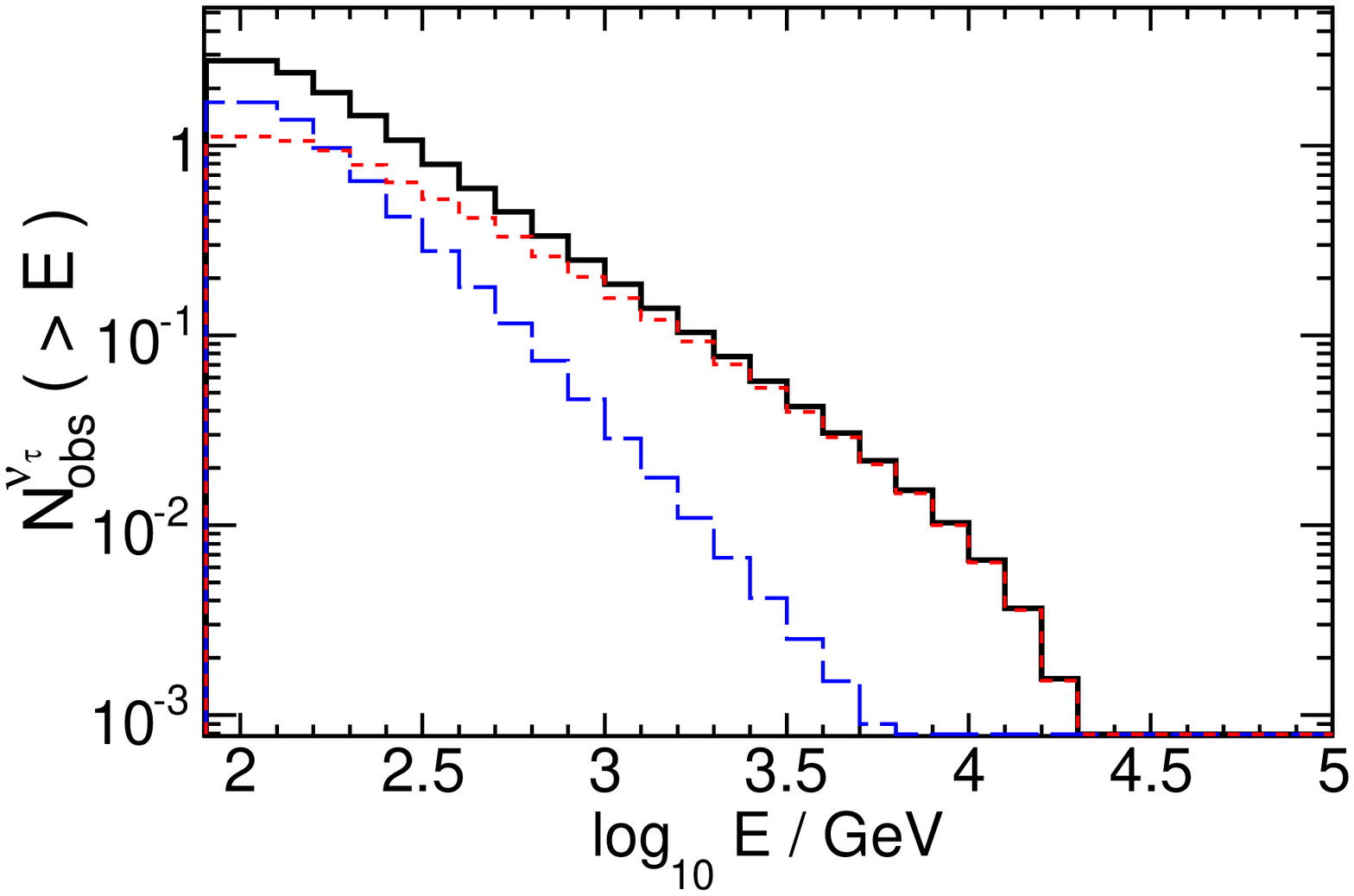}
\caption{\label{fig:ICrates} The panels show the expected signal
  above a given energy for cascades in IceCube/KM3Net for  $\nu_e$
  (left), $\nu_\mu$ (center) and $\nu_\tau$ (right). The solid line
  (black in the online version) is the total expectation of the RMW/AB
  model, the long dashed line (blue in the online version) is the pion
  contribution and the short dashed line (red in the online version)
  is the kaon contribution.}
\end{figure*}

The all-flavor expectation for DeepCore is 4 events. The
neutrino-induced cascade signal expected in IceCube is 6.1
events. DeepCore events are broken down by flavor to 0.4 due to
$\nu_e$, 3 are due to $\nu_\mu$ and 0.6 due to  $\nu_\tau$. Cascade
events break down to 2 due to $\nu_e$, 1.3 due to $\nu_\mu$ (via
neutral current interactions) and 2.8 due to $\nu_\tau$.

Figure \ref{fig:DCrates} shows the observed event spectra for all
three neutrino flavors in DeepCore due to a reference supernova at 10
Mpc that follows the RMW/AB model. Figure \ref{fig:ICrates} is similar
but neutrino-induced cascades in a km$^3$ detector. It is clear that
the pion component 
makes the most significant contribution to the observations in
DeepCore. In contrast events expected in IceCube, both $\nu_\mu$ and
cascades, are mostly due to kaons with the exception of energies near
to the detector threshold. 

\section{Atmospheric Neutrino background}

In DeepCore $10^5$ atmospheric neutrinos of all flavors over 2$\pi$~sr
are expected  per year \cite{deepcore}. IceCube has an atmospheric
muon neutrino rate of $10^5$ events per year over 2$\pi$~sr
\cite{IceCube}. IceCube has a rate of $2\times 10^4$ atmospheric
neutrino-induced cascades per year over 4$\pi$~sr\cite{Michelangelo}.

When searching for neutrino transients, background is significantly
reduced by using a narrow time window and by pointing in a specific
direction. Km$^3$ neutrino detectors have angular resolutions of
$O(1^\circ)$ \footnote{KM3Net may achieve better angular resolution
  than this for higher energy. But at 100~GeV - 1~TeV, the angular
  resolution is dominated by the directional difference between the
  parent $\nu_\mu$ and the daughter muon.}. Due to the optical
properties of ice, IceCube is not expected to be able to reconstruct
the direction of cascades well, but it is anticipated that KM3Net will
be able to reconstruct cascades with $5^\circ$ resolution
\cite{CascadeDirection}. It is also expected that DeepCore will
reconstruct the direction of events. For charged current $\nu_\mu$
events above $\sim$10-20~GeV DeepCore will have a resolution of
10-15$^\circ$. For cascade-like events in DeepCore the angular
resolution is expected to be $\sim 30^\circ$ but the minimum energy at
which this is feasible is still unknown. DeepCore will also have the
ability to separate track events (from C.C. $\nu_\mu$ interactions)
from cascade-like events for energies greater than 10~GeV
\cite{DarrenGrant}. In our calculations below we assume, perhaps
pessimistically, an angular resolution of $\sim 30^\circ$ for all
flavors in DeepCore. The appropriate time search window is set by the
size of the star: $R_{*} / c_{\mathrm light}$ $\sim 100 
\mathrm{s})$. In fact the radii of progenitors just prior to core
collapse are known with poor detail. Candidate progenitor stars have
been identified for about a dozen of core collapse supernovae,
including SN1987A (type IIP, $R_{*}  \sim 10^{11}$~cm) \cite{SN1987A}
and SN2008D (type Ib, $R_{*} \sim 10^{11}$~cm) \cite{SN2008D}. But
SN1987A is unusual and type II progenitors are more frequently red
giants with $R_{*}$ as large as $\sim 10^{13}$~cm.

The rate of accidental atmospheric neutrino multiplets with $N$ events of channel $a$
(e.g. $>$100~GeV $\nu_\mu$) and $M$ (e.g DeepCore) events of channel $b$ is:

\begin{equation}
\label{eq:yearlyrate}
R_{N,M} = R_{a}^N R_{b}^M (\frac{\Omega_a}{2\pi})^{N-1} (\frac{\Omega_b}{2\pi})^{M} \frac{\Delta T^{M+N-1}}{(N-1)! M!},
\end{equation} 
where $R_a$ and $R_b$ are the atmospheric neutrino background rates
for channels $a$ and $b$, $\Omega_a$ and $\Omega_b$ are the angular
areas search areas and $\Delta T$ is the temporal search window. Note
that eq. \ref{eq:yearlyrate} is only valid for $N\ge1$ and $M\ge0$. In
the case that the neutrino multiplet with a single species, it is
necessary to set $M=0$.

The rate of accidentals should be matched to at most, what is feasible
to follow-up with robotic optical telescopes, about 50/year. The
resulting angular uncertainty of the multiband or multiflavor neutrino
multiplet must also match the field of view of robotic optical
telescopes, about $2^\circ \times 2^\circ$.  Using
eq. \ref{eq:yearlyrate} we find that optical follow-ups are possible for
2 or more $\gtrsim$100~GeV $\nu_\mu$ (which has already been
proposed), but also (with KM3Net) one $\gtrsim$100~GeV $\nu_\mu$ and
one $\gtrsim$100~GeV cascade. Two $\gtrsim$100~GeV cascades also
produces an accidental rate that might be appropriate, but its follow
up would require a very large field of view telescope. Coincidences of
one $\gtrsim$100~GeV $\nu_\mu$ and at least three DeepCore events also
provides a good match to perform optical follow up. Tables
\ref{tab:table1} and \ref{tab:table2} show the accidental false rate
of neutrino multiplets for $\nu_\mu$, DeepCore events and cascades.

\section{Discussion}

Following the RMW/AB model, we have calculated the event expectation
in DeepCore and for cascades in km$^3$ detectors like IceCube and
KM3Net. We find that for a reference core-collapse supernova at 10 Mpc
$\sim 4$ events would be detected by DeepCore and $\sim~6$ cascades
would be detected by km$^3$ detectors. These signals are strong enough
to allow for the search of neutrinos in coincidence with a known
supernova in the scale of 10 Mpc. In the case of neutrino only 
searches, the atmospheric neutrino background is higher than what we
described in the text, as the time of the explosion can only be
established with about 1 day precision \cite{CowenKowalski}. 

A better alternative is to expand the already running optical
follow-up programs. A single TeV $\nu_\mu$ event can provide accurate
location in the sky, while a coincident observation with at least three
10~GeV events in DeepCore would result in an acceptable false
accidental rate. One muon event and one cascade event at 
$\gtrsim$100~GeV also has a very low accidental rate in KM3Net,
because of its good angular resolution for cascades. A DeepCore-like
component in KM3Net may have better angular resolution 
than in IceCube, but the level of improvement is limited by the
kinematical direction difference between the neutrino and the outgoing
muon or shower. 

The $\gtrsim$10~GeV observations provide another advantage. Because
they are sensitive to the pion contribution and the $\gtrsim$100~GeV
neutrinos are sensitive to the kaon contribution, the multiband
channel helps the IceCube/DeepCore combination to maintain sensitivity
if actual supernovae deviate from the reference model.

Observations of $\gtrsim$10~GeV and $\gtrsim$100~GeV neutrinos in
coincidence with core collapse supernovae would be very strong evidence
for the existence of choked jets. This observations would help
understand the correlation between long duration gamma ray bursts and
core collapse supernovae. Finally these observations may provide an
alternative way of detecting gamma-ray dark core collapse supernovae
with mildly relativistic jets as those observed in SN2007gr and SN2009bb.

The expansion of the optical follow-up programs proposed here is
promising in light of the core-collapse supernova rate within 10 Mpc:
1-2  core collapse SNe/yr \cite{Kistler:2008us,Ando:2005ka}. The expected opening
angle of the choked jets implies a random aligning of one of the jets with the
line of sight for 10-20\%. This would result in a positive
observation every 2.5 - 10 years within 10 Mpc if all core collapse
supernovae have hidden jets. But this is a conservative estimation of
the relevant supernova rate. The neutrino event rates calculated in this letter imply that
observations are possible at distances beyond 10 Mpc and the supernova
rate depends of the cube of the distance \footnote{At close distances
  the detailed distribution of local galaxies is important and does
  deviate somewhat from the continuous limit}.

\begin{table}
\caption{\label{tab:table1} Expected coincident yearly rates of N $\nu_\mu$
  events and M DeepCore events. We assume the $\nu_\mu$ events
  coincident in a circle of 1$^\circ$ radius and 30$^\circ$ for DeepCore
  events. We assume coincidence time window of 100~s. } 
\begin{ruledtabular}
  \begin{tabular}{ | c | | c | c | c | c | }
   DC /\ $\nu_\mu$  & 0 & 1 & 2 & 3 \\
    \hline
    0 & -      & -                              & 4.8     & 1.1$\times 10^{-4}$ \\
    1 & -      & -                              & 0.2     & 5.0$\times 10^{-6}$ \\
    2 & -      & 94                            & 4.6$\times 10^{-3}$ & 1.1$\times 10^{-3}$ \\
    3 & 94    & 1.4                           & 6.6$\times 10^{-5}$ & 1.6$\times 10^{-9}$ \\
    4 & 1.4   & 1.5$\times 10^{-2}$ & 7.2$\times 10^{-7}$ & 1.7$\times 10^{-11}$ \\
  \hline
  \end{tabular}
 \end{ruledtabular}
\end{table}

\begin{table}
\caption{\label{tab:table2} Expected coincident yearly rates of N $\nu_\mu$
  events and M cascade events. We assume the $\nu_\mu$ events
  coincident in a circle of 1$^\circ$ radius and 5$^\circ$ for cascade 
  events. We assume coincidence time window of 100~s.}
\begin{ruledtabular}
  \begin{tabular}{ | c | | c | c | c | c | }
    casc /\ $\nu_\mu$   & 0 & 1 & 2 & 3 \\
    \hline
    0 & - &  - & 4.8 & 1.1$\times 10^{-4}$ \\
    1 & - & 12 & 5.8$\times 10^{-4}$ & 1.4$\times 10^{-8}$ \\
    2 & 1.2 & 7.3$\times 10^{-3}$ & 3.5$\times 10^{-8}$ & 8.5$\times 10^{-13}$ \\
    3 & 7.3$\times 10^{-5}$ & 2.9$\times 10^{-8}$ & 1.4$\times 10^{-12}$ & 3.4$\times 10^{-17}$\\
  \hline
  \end{tabular}
 \end{ruledtabular}
\end{table}

\begin{acknowledgments}
We are grateful to John Beacom and Francis Halzen for helpful comments and
discussion. We are grateful to Darren Grant for providing information regarding
the performance of DeepCore. We thank the anonymous referee for useful
comments. This work has been partially supported by
National Science Foundation grant PHY-0855291. 
\end{acknowledgments}

\end{document}